%% file: HTC_framework.tex
\documentclass[conference]{IEEEtran}

\pdfoutput=1 
\usepackage[usenames, dvipsnames]{color}
\usepackage{multirow}
\usepackage{booktabs}
\usepackage{amsmath}
\usepackage{amsfonts}
\usepackage{multirow}
\usepackage{epsfig}
\usepackage{psfrag}
\usepackage{pifont}
\usepackage{cite}
\usepackage{url}
\usepackage{algorithm}
\usepackage{algorithmic}
\usepackage{graphicx}
\graphicspath{ {./figs/} }
\usepackage[usenames]{color}
\usepackage{amssymb,amsmath}
\usepackage{amsthm}
\usepackage{subcaption}
\usepackage{textcomp}
\usepackage{wrapfig}
\usepackage{floatflt}
\usepackage[utf8]{inputenc}
\usepackage{tabularx}
\usepackage{subcaption}

\usepackage{amsmath}
\usepackage{amsfonts}

\newcounter{numberlistc}

\newcounter{itemlistc}

\newenvironment{itemlist}
    {   \setcounter{itemlistc}{0}
    \begin{list}{$\bullet$}
        {\usecounter{itemlistc}
        \setlength{\parsep}{0pt}
        \setlength{\topsep}{3pt}
        \setlength{\itemsep}{0pt}}
        }{ \end{list} }

\def\BibTeX{{\rm B\kern-.05em{\sc i\kern-.025em b}\kern-.08em
    T\kern-.1667em\lower.7ex\hbox{E}\kern-.125emX}}
 
\small
\usepackage{array}
\newcolumntype{P}[1]{>{\centering\arraybackslash}p{#1}}
\begin{document}

\title{Hybrid Temporal Computing for Lower Power Hardware Accelerators}

\author{ \IEEEauthorblockN{Maliha Tasnim, Sachin Sachdeva,  Yibo Liu and Sheldon X.-D. Tan
    \thanks{The work is supported in part by NSF grant under
      No.CCF-2007135, and in part by NSF grant under No. CCF-2113928. Corresponding author: Sheldon Tan at stan@ece.ucr.edu}
  } \IEEEauthorblockA{
    Department of Electrical and Computer Engineering, University of California, Riverside, CA 92521 USA}}

\maketitle

\begin{abstract}


In this paper, we propose a new hybrid temporal computing (HTC) framework that leverages both pulse rate and temporal data encoding to design ultra-low energy hardware accelerators. Our approach is inspired by the recently proposed temporal computing, or race logic, which encodes data values as single delays, leading to significantly lower energy consumption due to minimized signal switching. However, race logic is limited in its applications due to inherent restrictions. The new HTC framework overcomes these limitations by encoding signals in both temporal and pulse rate formats for multiplication and in temporal format for propagation. This approach maintains reduced switch energy while being general enough to implement a wide range of arithmetic operations. We demonstrate how HTC multiplication is performed for both unipolar and bipolar data encoding and present the basic designs for multipliers, adders, and MAC units. Additionally, we implement two hardware accelerators: a Finite Impulse Response (FIR) filter and a Discrete Cosine Transform (DCT)/iDCT engine for image compression and DSP applications. 
Experimental results show that the HTC MAC has a significantly smaller power and area footprint compared to the Unary MAC design and is orders of magnitude faster. Compared to the CBSC MAC, the HTC MAC reduces power consumption by $45.2\%$ and area footprint by $50.13\%$. For the FIR design, the HTC design significantly outperforms the Unary design on all metrics. Compared to the CBSC design, the HTC-based FIR filter reduces power consumption by $36.61\%$ and area cost by $45.85\%$. The HTC-based DCT filter retains the quality of the original image with a decent PSNR, while consuming $23.34\%$ less power and occupying $18.20\%$ less area than the CBSC MAC-based DCT filter.
\end{abstract}

\input Introduction.tex

\input Literature_Review.tex
\input new_temp_design.tex

\input Temporal_SC_MAC.tex
\input Experimental_Result.tex

\input Conclusion.tex

\bibliographystyle{ieeetr}
\bibliographystyle{ieeetr}
\bibliography{./ref.bib,./ucr.bib}
\end{document}

%% file: Introduction.tex
\section{Introduction}

One of the paramount challenges confronting today's computing landscape is the exponential surge in computing power due to emerging generative AIs contrasted with the linear progression of power supply. This illustration underscores the projection that computing energy consumption will roughly double every three years, while global energy production is experiencing only a modest linear growth rate of approximately $2\%$ per year~\cite{decadal_plan}. Mitigating this exponential growth requires substantial advancements in computing energy efficiency.

To significantly reduce energy consumption, fundamentally new and ultra-low energy computing paradigms are urgently needed. One such paradigm is {\it temporal computing}~\cite{Madhavan:ISCA'14,Madhavan:JETCS'21}, which is rooted in the concept of race logic~\cite{Madhavan:ISCA'14}. In race logic, information is not encoded in individual bits, but rather in the timing of voltage transitions from low to high. This method allows multiple bits of information to be encoded on a single wire. Although this approach sacrifices precision compared to traditional binary representation, it offers unique advantages in speed, energy efficiency, and reduced area footprint. However, race logic-based temporal computing still has limited applications, as it relies on wavefront travel, which is more suited for problems that can be mapped to spatially organized data structures such as graphs or trees. Examples include DNA global sequence alignment~\cite{Madhavan:ISCA'14}, decision tree-based classification~\cite{Tzimpragos:ASPLOS'19}, and sorting networks~\cite{Smith:ISCA'18}. Leveraging temporal computing for general-purpose computing remains a challenging problem.


Meanwhile, a promising avenue for achieving significant energy reduction is approximate computing~\cite{Han:ETS'13,Venkataramani:DAC'15}. One prominent method in this domain is stochastic computing (SC), where values are represented as probabilities in a bit stream or pulse rate rather than conventional binary numbers~\cite{AlaghiQian:TCAD'18}. SC offers an inherent progressive trade-off between accuracy and latency/energy/area by adjusting the length of bit streams, making it extremely low-cost and energy-efficient. However, traditional SC hardware implementations suffer from very long latency and large area overhead due to the need for random number generation.

Recently, a more
efficient and also more accurate SC multiplier, called counting-based SC or CBSC,  was proposed to
partially mitigate the two aforementioned problems in traditional
SC~\cite{SimLee:DAC'17}. First, It replaces the {\it AND} operation with a `1' counting process in the bit stream, which can be reduced without going through the entire length of the bit streams, and second, the bit streams no longer need to be random. Many follow-up works have improved this counting-based SC scheme~\cite{Chen:ICCAD'20,Yu:DAC'21,Yu:DAC'22}. But those methods still can't directly perform the addition in bitstream as the multiplication results are translated into binary after the counting process. On the other hand, several other deterministic SC methods were proposed recently~\cite{Jenson:ICCAD'2016,Najafi:TVLSI'2019,Najafi:journal'2017,Schober:IEEE'21}. However, these methods either suffer from long latency issues~\cite{Jenson:ICCAD'2016,Najafi:TVLSI'2019} or require special analog circuits for the conversion of time or delay to binary data. 


In this article, we propose a novel hybrid temporal computing framework that leverages both bitstream (pulse rate) and temporal data encoding to design ultra-low energy hardware accelerators. This approach can also be viewed as a generalized CBSC framework~\cite{SimLee:DAC'17,Yu:DAC'21}, enabling addition operations to be performed in the bitstream format. The key contributions of this work are summarized below:

\begin{itemlist}


 \item First, we propose a general hybrid temporal computing (HTC) framework and associated arithmetic operations based on a uniform bitstream format. Unlike existing stochastic computing, our approach encodes data in two formats: temporal (e.g., delay or a single pulse) and traditional bitstream or pulse rate format. This design minimizes switching activities while retaining the energy efficiency of stochastic computing. Further more,  the temporal data format further reduces energy consumption for  signal propagation between computing. 
 

 \item In the HTC framework, all bitstreams are generated deterministically, avoiding the need for costly random number generation circuits. Multiplication is performed using one temporal data and one regulated bitstream data with an AND operation, which can be viewed as a special counting process in the CBSC method, offering the same accuracy~\cite{SimLee:DAC'17, Yu:DAC'21}. We demonstrate how these operations can be executed for both unipolar and bipolar uniform encoding. Addition is performed using traditional scaled addition in SC with a MUX gate, which can be further improved using more accurate SC-like addition. Compared to traditional SC, the proposed method has a significantly smaller hardware footprint while achieving the same or even better accuracy and reduced energy consumption.

\item Compared to existing deterministic SC methods~\cite{Jenson:ICCAD'2016,Najafi:TVLSI'2019,Najafi:journal'2017,Schober:IEEE'21}, the proposed method does not increase bitstream to ensure the accuracy~\cite{Jenson:ICCAD'2016,Najafi:TVLSI'2019} and does not requires special analog circuit to convert time/delay signal back to binary data~\cite{Najafi:journal'2017,Schober:IEEE'21}. The temporal data can be naturally converted back to binary with a counter or shift register. 

\item We show the architectures of multiplier-accumulator (MAC)  for the proposed HTC design and one specific 4-input MAC design. The design features hybrid data inputs (temporal data and general bitstream format) and the temporal data output to further reduce energy for data propagation due to minimized switching activities. 

\item Experimental results show that the HTC MAC uses only a small marginal power and area footprint compared to the Unary MAC design. It is also orders of magnitude faster than the Unary design. Compared to the CBSC MAC, it reduces power consumption by $45.2\%$ and area footprint by $50.13\%$.

\item Furthermore, we implemented two hardware accelerators: a Finite Impulse Response (FIR) filter and a Discrete Cosine Transform (DCT)/iDCT engine for image and DSP applications. HTC designs  again significantly outperform the Unary design across all metrics. Compared to CBSC design, the HTC MAC-based FIR filter reduces power consumption by $36.61\%$ and area cost by $45.85\%$.  The HTC-based DCT filter retains the quality of the original image with a decent PSNR, while consuming $23.34\%$ less power and occupying $18.20\%$ less area than the state-of-the-art CBSC MAC-based DCT filter.

\end{itemlist}

This paper is structured as follows: Section~\ref{sec:rela_work} reviews temporal logic, recent advances in deterministic stochastic computing, counter-based stochastic computing, and related work. Section~\ref{sec:architecture} introduces the proposed concept of hybrid temporal computing, the design of basic building blocks such as adders and multipliers, and the specific design of a 4-input MAC. Section~\ref{sec:results} presents the experimental results. Finally, Section~\ref{sec:conclusion} concludes the paper.

%% file: Literature_Review.tex
\section{Preliminary and review of related works}
\label{sec:rela_work}

\subsection{Review of temporal computing and race logic}

\begin{figure}
\centering
\includegraphics[width=0.95\columnwidth]{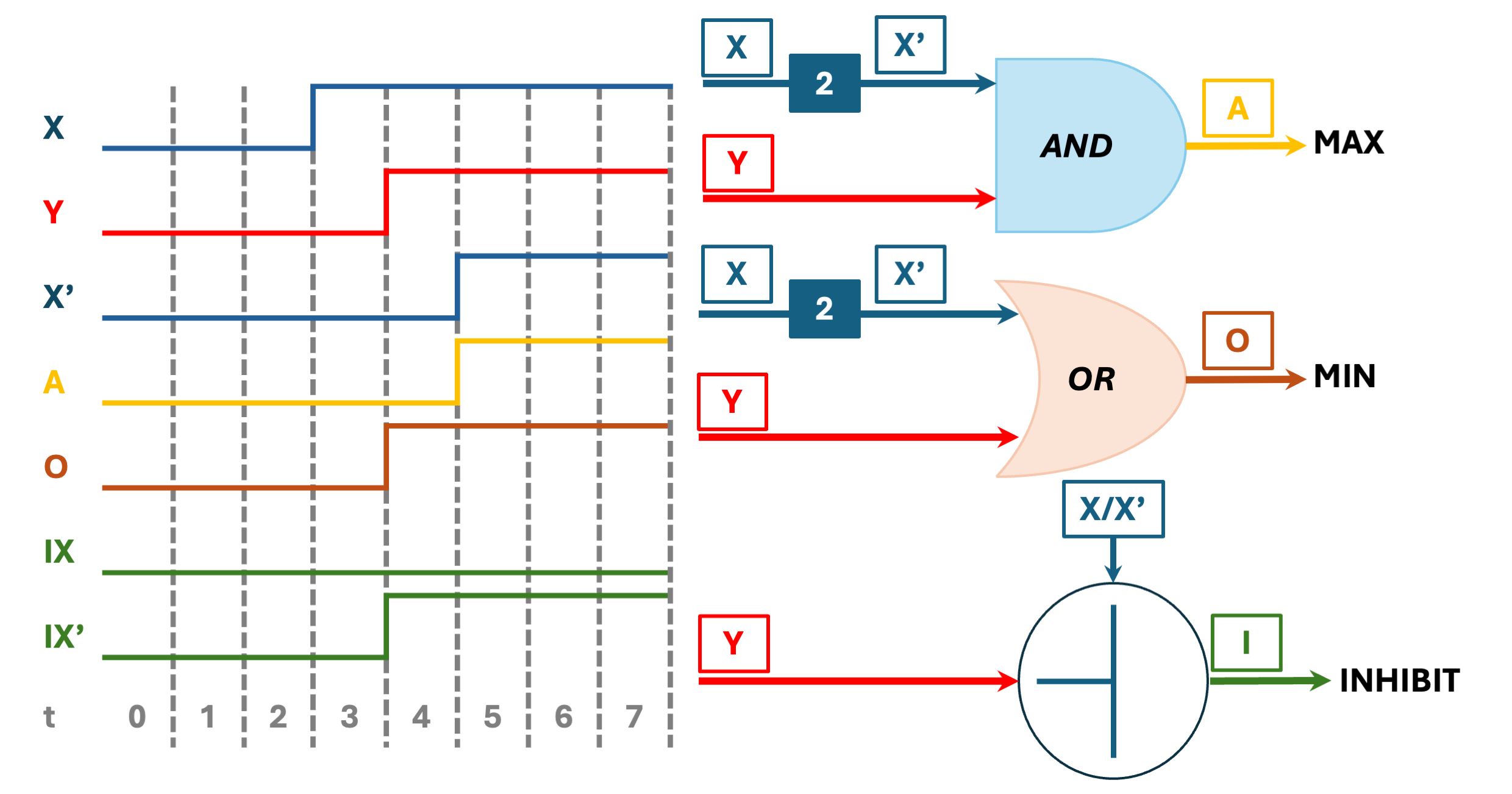}
\caption{As illustrated, in race logic, X indicates 2 (2 delay from the origin time), Y=3, X'=4, O is min (X' and Y) which is 3. A is max of (X' and Y), which is 4. INHIBIT means if Y arrives earlier than X/X', then I=Y, otherwise I = unchanged~\cite{temporal_computing_nist}.}
\label{fig:race_logic_ill}
\vspace{-0.1in}
\end{figure}

One effective method for reducing energy consumption is to alter the signal encoding. {\it Temporal computing}, recently proposed in the literature~\cite{Madhavan:ISCA'14,Madhavan:JETCS'21}, is based on the concept of race logic \cite{Madhavan:ISCA'14}. Race logic uses a relative timing code to represent information through a wavefront of digital edges on wires. Initially proposed to accelerate dynamic programming and machine learning algorithms, such as DNA global sequence alignment \cite{Madhavan:ISCA'14}, race logic encodes information based on the relative timing between rising edges on different wires with respect to a temporal reference. This encoding scheme allows each wire to represent multiple bits of information based on the arrival time of the rising edge, while maintaining low activity factors due to the single transition per wire during computation. Consequently, conventional Boolean primitives can perform non-traditional operations with minimal energy costs. For example, a single OR gate can effectively execute a two-input MIN function, as depicted in Fig. \ref{fig:race_logic_ill}. Other fundamental race logic primitives such as MAX, INHIBIT, and ADD-BY-CONSTANT gates, shown in the figure, have been demonstrated to be sufficient for implementing any temporally invariant and causal functions within the proposed space-time algebra \cite{Smith:ISCA'18}.


Race logic demonstrates orders-of-magnitude improvements in energy efficiency compared to classical approaches~\cite{Madhavan:JETCS'21}. This efficiency is attributed to its minimal power switch activities in the wires, where signals switch only once per algorithm process, and the progression of physical time incurs no additional cost. While race logic is constrained by the limited precision inherent in analog-based computing, studies have shown that for neocortical computing, 3-4 bit accuracy is sufficient \cite{Yamazaki:BrainSci'22}. It is important to differentiate race logic from another bio-inspired approach, Spiking Neural Networks (SNN) \cite{Yamazaki:BrainSci'22}, which encode signals in terms of spike rates or the number of spikes, whereas race logic encodes signals based on exact spike timing. Both SNN and race logic exploit temporal coding to enhance energy efficiency and can be conceptualized within the framework of temporal neural networks (TNN)~\cite{Smith:ISCA'18}.

Race logic design techniques have found diverse applications, from accelerating dynamic programming algorithms \cite{Madhavan:ISCA'14} to DNA global sequence alignment \cite{Madhavan:ISCA'14}, decision tree-based classification \cite{Tzimpragos:ASPLOS'19}, and sorting networks \cite{Smith:ISCA'18}. It effectively tackles problems with graph-tracking properties like shortest path searching and sorting by computing the timing of edge wavefronts. However, pure temporal computing faces challenges in digital signal process (DSP) and  deep neural networks due to its inability to perform general arithmetic operations like multiplication and additions, constrained by causality and waveform format restriction. Despite recent efforts to develop temporal state machines \cite{Madhavan:JETCS'21} and temporal memory structures \cite{Madhavan:ISCAS'20, Vakili:JESCDC'20, Madhavan:ISCAS'20}, race logic-based design for general computing still remain challenging. 

On the other hand, recently, both temporal and rate coding have been applied in the design of DSP and CNN networks, utilizing superconducting devices~\cite{Gonzalez-Guerrero:ASPLOS'22,Gonzalez-Guerrero:ISQED'23}. These approaches leverage the generation of spikes or pulses, known as Single Quantum Flux (SQF), in Josephson junctions, crucial components in superconductor-based designs. 
However, this design has primarily been explored for superconductor devices, where pulses are naturally generated from Josephson junctions. Nonetheless, superconductor-based computing faces challenges, including a $1000\times$ device density gap compared to existing CMOS designs, as well as the requirement for low temperatures.

\subsection{Review of deterministic stochastic computing}
\label{sec:review_of_sc}

\begin{figure}
    \centering
    \includegraphics[width=0.49\textwidth]{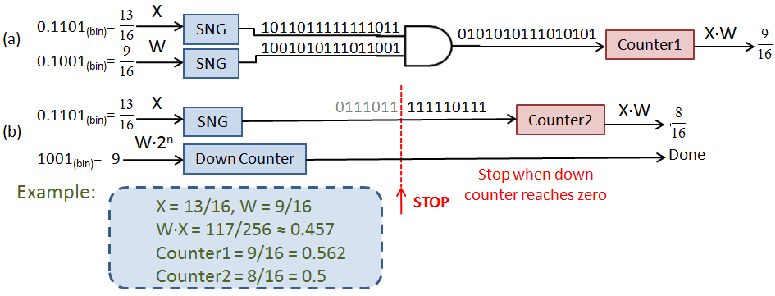}
 \caption{(a) Traditional stochastic multiplication; (b) The counting-based stochastic (CBSC)  multiplication~\cite{SimLee:DAC'17,Yu:DAC'21}}
    \label{fig:cbsc_mul_concept}
    \vspace{-0.1in}
 \end{figure}

Stochastic computing (SC) can be conceptualized as a specialized form of temporal computing, where binary data is transformed into sequential bitstreams. In SC, the multiplication of two bitstreams is achieved through simple AND operations for unipolar coding, analogous to the MAX operation in temporal logic, as illustrated in Fig.~\ref{fig:cbsc_mul_concept} (a). 

Recent advancements in counting-based SC (CBSC) multiplication have eliminated the need for bitstream randomness without compromising accuracy~\cite{SimLee:DAC'17,Yu:DAC'21}. 

This approach, illustrated in Fig.~\ref{fig:cbsc_mul_concept} (b), replaces the AND operation with a deterministic counting process. Here, the number of ones from input $X$ is decremented by a timer based on the values of input $W$. 
Different from the traditional SC multiplication, CBSC only requires one FSM (finite state machine)-based SNG to convert one of the two binary inputs, ex. $X$, into a bitstream
with deterministic pattern first.  The FSM-based SNG (stochastic number generator) evenly distributes the
$X_{i-1}$, which is the $i$th bit of $X$, based on its binary weight
$2^{i-1}$.  For instance, if $i=3$, then $X_2$ will appear 4 times in
the resulting SN. Such SN generation can be simplified and implemented
by an FSM and a MUX. The FSM is actually an up counter counts from 0
to $2^N-1$, assuming $X$ is $N$-bit.  The MUX then outputs $X_{i-1}$
based on the output value of the FSM.

If the SN bitstream for the other input $W$ is set to be series of
`1' followed by a bunch of `0' as shown in Fig.~\ref{fig:cbsc_mul_concept}.  As SC
multiplication is simply {\it AND} operation, it is not necessary to
count the second half of the output bitstream.  So, the whole
counting process only requires $W \cdot 2^N$ cycles to finish, which
saves half latency in average. The authors used a down counter to
realize the idea. While $w \cdot 2^N$ is used as the initial value,
the down counter decreases by one in each clock cycle. When it reaches
``zero'', the process is terminated. As a result, CBSC leads to a
simpler design as one traditional SNG (typically using LFSR) and the {\it
  AND} gate are removed in exchange of a down counter, which is
much cheaper than an SNG.

CBSC has evolved beyond stochastic computing, as it no longer relies on stochastic processes. Instead, it should be regarded as a specialized form of temporal computing, laying the foundation for the proposed hybrid temporal computing paradigm, as we will elaborate in subsequent sections.  Recent advancements in CBSC have further enhanced its capabilities~\cite{Chen:ICCAD'20,Yu:DAC'21,Yu:DAC'22}, including improvements in counting techniques~\cite{Yu:DAC'21}, scaled implementations for enhanced accuracy~\cite{Yu:DAC'22}, and the design of more accurate CBSC multipliers (scaled CBSC)~\cite{Yu:DAC'22}. However, this method has yet to be extended to develop complete arithmetic computing units such as adders and multiplier-accumulators (MACs) and processing units, as the counting process typically operates in binary format (originally termed binary interface SC)~\cite{SimLee:DAC'17}.

In addition to CBSC, several deterministic SC computing designs have been proposed recently. The early work for performing SC using deterministic bitstream was proposed in~\cite{Jenson:ICCAD'2016,Najafi:TVLSI'2019}. The idea is to use repeat bitstreams of one operands for each bit of other operand. However, such setup will leads to quadratic bit stream increase over each operation. To mitigate this problem, bitstream lenth reduction and approximation was proposed later~\cite{Kiran:GLSVLSI'22}. 


Recently a time-encoded stochastic computing has been proposed~\cite{Najafi:journal'2017}, which uses analog-like pulse width modulation (PWM) to represent the signal value (the duty cycles of the pulses). This method is called {\it Unary design} in this work for comparison propose. 
As PWM is also uniform encoding (like SC), the multiplication (using AND or XOR) and addition (scaled addition) can be performed similarly to SC computing. However, since the bitstream is deterministic, the operands have to be repeated many times (least common multiple (LCM) of the two operands). Since the final signal is in the PWM format, some time-to-voltage analog conversion has to be performed. To further improve accuracy, the exact addition for PWM-based computing was proposed where a software loop is introduced to produce exact delay (duty cycles) and non-overlapping pulses of two operands before the OR operation~\cite{Schober:IEEE'21}. 

For example, for two numbers $A$ and $B$ to multiply and $n_A$ and $n_B$ are the size of the resulting bitstreams for $A$ and $B$.  if the two bit streams of relatively prime lengths are considered, an effective strategy involves repeating both bit stream patterns for LCM (least common multiple) $(n_A,n_B)$ times. Additionally, introducing $N$ unique delays is essential for obtaining accurate results when summing with an OR gate. The accuracy of the the Unary design in theory can be accurate as binary design, but practically the accuracy of these results is contingent upon the number of ones allowed in the input bit streams, denoted as $v$, described in ~\cite{Schober:IEEE'21}. According to the method discussed, the results remain accurate provided that $v \leq \left\lfloor \frac{n}{\left\lceil \frac{\sqrt{4N+1}-1}{2} \right\rceil + 1} \right\rfloor$. This condition illustrates a boundary on the capability of the method to handle higher densities of ones in the input streams. Furthermore as shown in the result section, we found the performance metrics of the Unary-based design are significantly less than that of the proposed HTC and CBSC designs. 

%% file: new_temp_design.tex
\section{The new circuit level design for new hybrid temporal logic}
\label{sec:architecture} 
In this section, we begin by detailing the design of circuit and architecture-level components for fundamental arithmetic computing units. These include multipliers, accumulators, and multiplier-accumulators (MACs), all tailored for the proposed hybrid computing framework.

\subsection{Multiplication and addition in HTC}
\begin{figure}[ht!]
    \vspace{-0.2in}
     \includegraphics[width=0.49\textwidth]{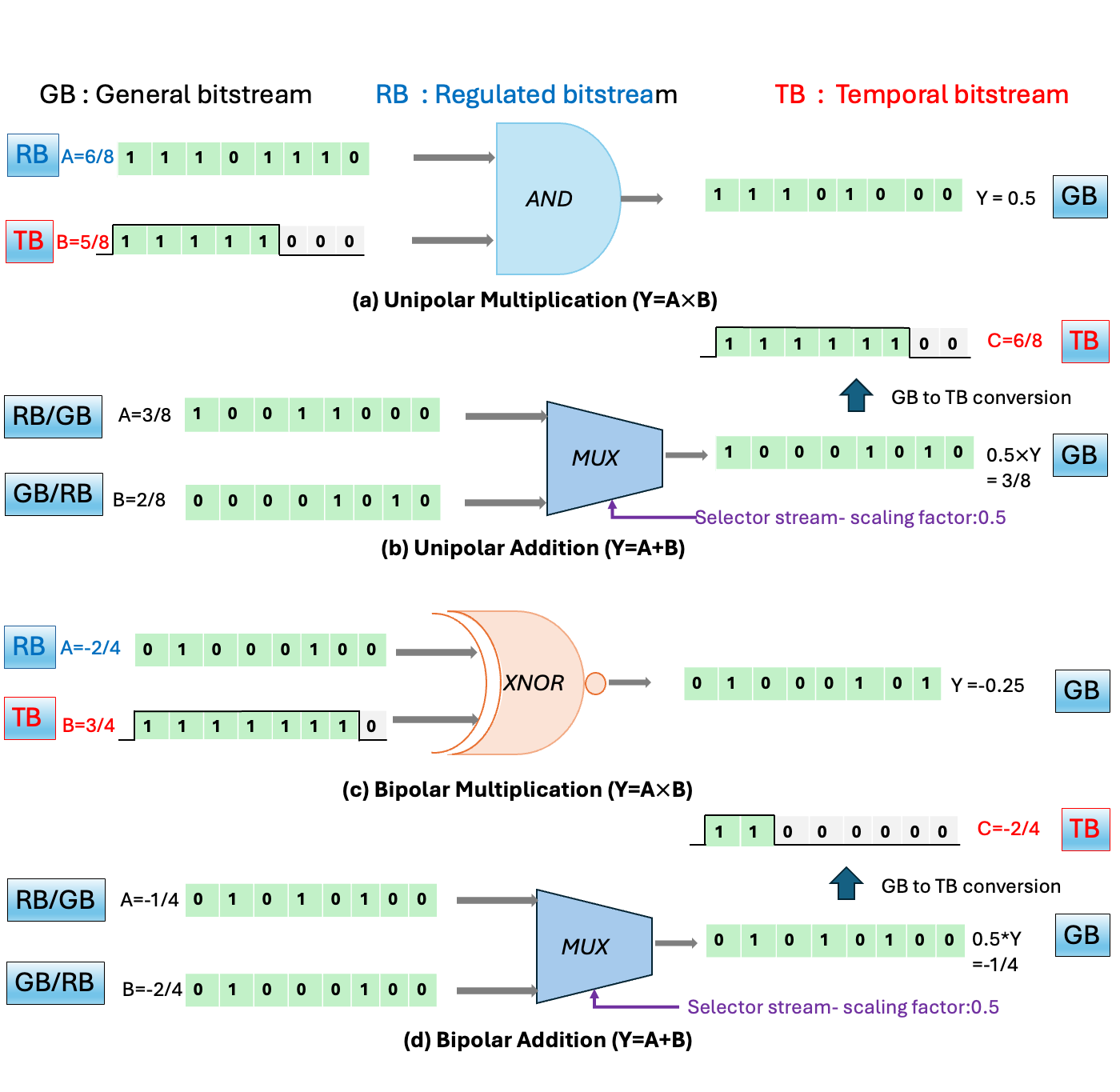}
 \caption{The multiplication and addition in HTC framework}
  \label{fig:multiply_and_add_htc} 
\end{figure}

In the Hybrid Temporal Computing (HTC) framework, data can be represented in two distinct formats, as demonstrated in Fig.~\ref{fig:multiply_and_add_htc}: a {\it General Bitstream} (GB) format, where the value is indicated by the number of '1' bits, and a {\it Temporal Bitstream} (TB) format, where the value is represented by the time period or delay relative to a reference signal. These two data patterns are illustrated in Fig.~\ref{fig:multiply_and_add_htc}. Typically, data values lie within the range of $[0,1]$ for unipolar coding and within the range of [-1,1] for bipolar coding.


In unipolar format, a real-valued number \( X \) is defined as \( X = p = \frac{n}{N_{\text{max}}} \), where \( n \) represents the number of '1' bits and \( N_{\text{max}} \) is the maximum '1' count or the number of clock cycles within the computing period (referred to as the epoch). For TB data \( p \), \( n \) denotes the count of '1' bits within that specific timing period. In the bipolar format, the data value is interpreted as \( X = 2p - 1 \) for the range of [-1,1], albeit with half the precision of the unipolar coding.

\begin{figure}[h]
   \vspace{-0.2in}
    \centering
    \label{fig:rb_data_example}
    \begin{subfigure}{0.45\columnwidth}
    \includegraphics[width=0.9\textwidth]{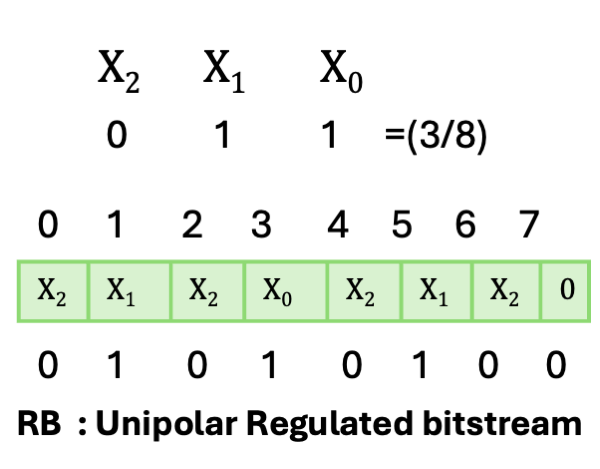}
     \caption{}
     \label{fig:unipolar_rb_data_example}
    \end{subfigure}
    \begin{subfigure}{0.45\columnwidth}
        \includegraphics[width=0.9\textwidth]{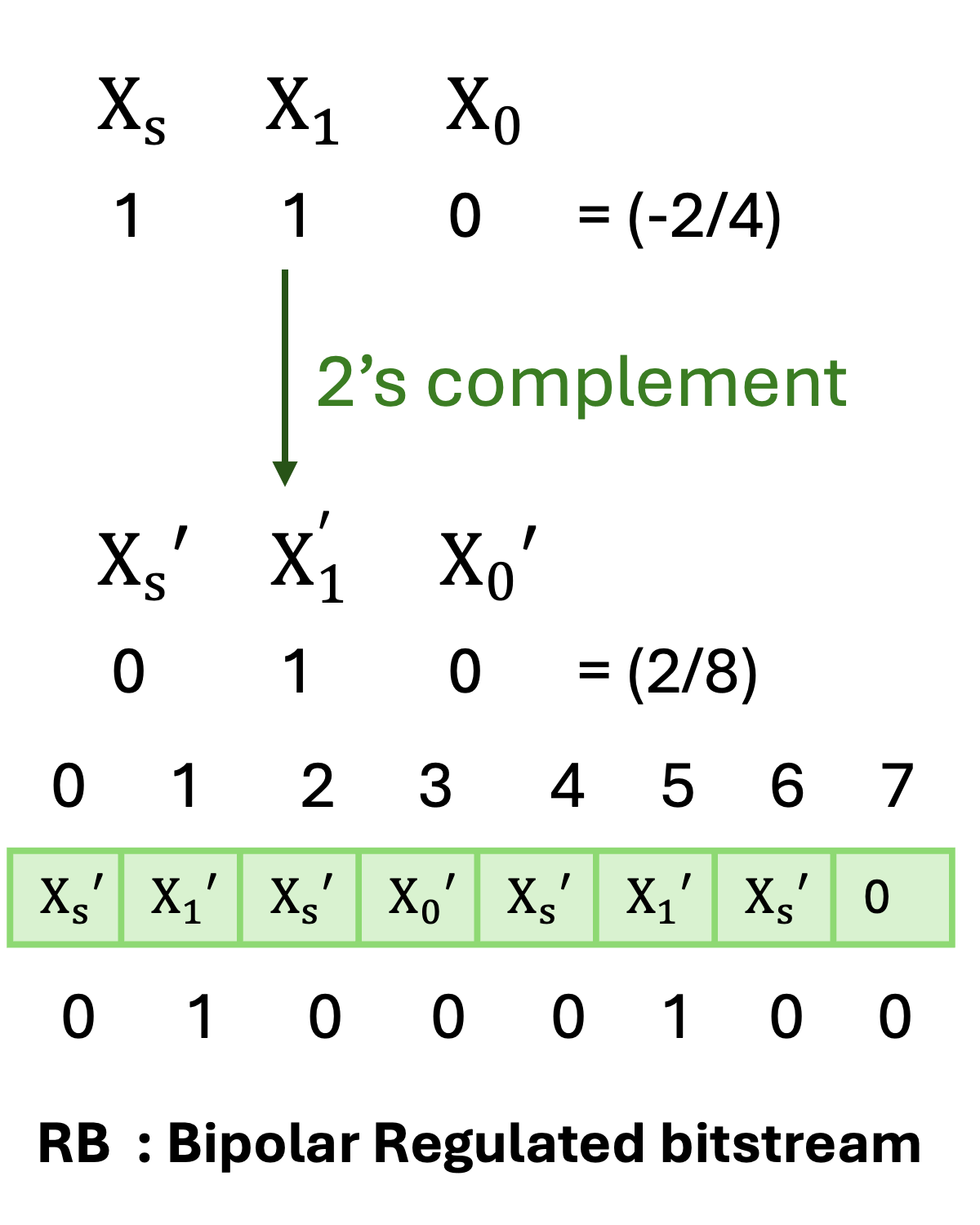}
     \caption{}  
         \label{fig:bipolar_rb_data_example}
    \end{subfigure}
    \caption{The regulated bitstream representation of (a) 3 bit unipolar binary data 011 and (b) 3 bit bipolar binary data 110. Here, $X_s$ is the sign bit. }
\end{figure}

Furthermore, data can be represented using the so-called {\it Regulated Bitstream} (RB), as illustrated in Fig.~\ref{fig:unipolar_rb_data_example}. This RB representation is similar to the bitstream utilized in CBSC~\cite{SimLee:DAC'17,Yu:DAC'21}. The concept involves {\it evenly} distributing each bit in the binary data across its corresponding bitstream, which is essential for HTC multiplication. Such regulated bitstreams can be generated using a straightforward finite-state machine.

The bipolar regulated bitstream in HTC is generated similarly to its unipolar counterpart using a finite state machine, as shown in Fig.~\ref{fig:bipolar_rb_data_example}. Let's consider a 3-bit signed binary number 110, which corresponds to \(-2\) or the scaled number \(-2/4\). To generate the RB for 110, we first compute the required number of '1's for an 8-bit stream as \(p = (X+1)/2 = (-2/4 +1)/2 = 2/8\). Thus, we need 2 '1' bits, represented as 010 in binary. Subsequently, we generate the bitstream pattern \(0100010\) using the same method for unipolar coding, as shown in Fig.~\ref{fig:bipolar_rb_data_example}. In our implementation, we generate 010 by performing bit-level operations, specifically the 2's complement operation on the signed binary number 110.

Using a similar method, we can generate the bitstream for the TB format for all bipolar numbers. The process remains largely the same, with the distinction that once we compute the number of '1's needed, all these '1's are consecutively placed to form the TB bitstream. For instance, in Fig. \ref{fig:multiply_and_add_htc} (c), if we intend to generate a TB for \(3/4\), we calculate the number of '1' bits out of 8 as \(p = (3/4 + 1)/2 = 7/8\). Consequently, we can generate \(11111110\) as the TB stream for \(3/4\).



In this work, we focus on computationally intensive accelerators like DSPs and deep neural networks, inputs often undergo multiplication by coefficients or weights. Since these coefficients are typically loaded into memory once and rarely updated thereafter, they can be efficiently represented using the GB format.

Next, we introduce our HTC multiplication, addition/accumulator, and process element unit. Unlike existing CBSC-based computing methods that rely on counting, our approach utilizes a simple AND (XNOR) gate for unipolar (bipolar) coding to perform multiplication. This leverages uniform encoding of both RB and TB data, akin to SC bitstreams, allowing us to adapt multiplication and scaled additions into our HTC computing framework. For bipolar multiplication, the sign bit is incorporated into both the regulated bitstream and the 2's complemented temporal bitstream. Consequently, the output from the XNOR gate for multiplication is scaled by 0.5.


Fig. \ref{fig:multiply_and_add_htc} (first figure) illustrates HTC multiplication. In this example, input $A$ (110) represents $6/8$, a 3-bit positive number in RB format, while input $B$ is represented as $5/8$ in TB format. The multiplication is performed using an AND operation. The exact result $Y$ should be $0.469$, whereas HTC multiplication yields $0.5$, providing a close approximation.

HTC addition can effectively leverage traditional SC methods by employing scaled addition using a simple MUX gate with a selection probability of $0.5$, as depicted in Fig. \ref{fig:multiply_and_add_htc} (second figure). Thus, the sum of two numbers is twice the probability of '1' in the output bitstream of the MUX gate. After addition, the resultant bitstream is in the general bitstream format. To propagate the data for the next stage of computation, conversion to TB format is achieved using a straightforward shift register. When converting the final data back to binary, a counter can be employed at the end of the operation. Consequently, our method eliminates the need to convert TB signals back to binary, a requirement in existing deterministic SC methods~\cite{Jenson:ICCAD'2016,Najafi:TVLSI'2019}


To perform multiplication using bipolar encoding, it can be achieved using an 'XNOR' gate with one operand in GB format and the other in TB format, as illustrated in Fig. \ref{fig:multiply_and_add_htc} (third figure). In this example, the input regulated bitstream or GB $A$ represents $-2/4$ (110 in the signed binary domain), while the input temporal bitstream or TB $B$ represents $3/4$ (011 in binary domain).
Bipolar HTC addition follows a similar approach to unipolar addition, where a 2:1 mux is utilized for scaled addition, depicted in Fig. \ref{fig:multiply_and_add_htc} (fourth figure).


%% file: Temporal_SC_MAC.tex
\subsection{Hardware implementation of multiple-input HTC MAC}

\begin{figure}
    \centering
    \includegraphics[width=0.4\textwidth]{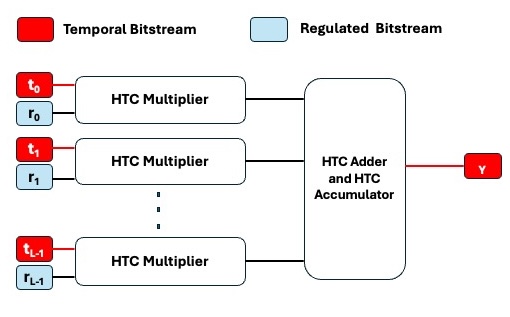}
 \caption{The proposed HTC multiple-input MAC architecture}
 \label{fig:htc_dot_product}
    \vspace{-0.1in}
\end{figure}

In this section we describe specific hardware implementation for multiplication and accumulation (MAC) operation.  MAC operation  essentially is to performed $a \leftarrow  a + (b \times c)$ operation. We  practically may have multiple multiplications and one multi-input summation (also called dot product) as following:
\begin{equation}
    a =  \sum_{i=1}^{N} ( b_i \times  c_i)
    \label{eq:mul_mac_op}
\end{equation}
Fig. \ref{fig:htc_dot_product} shows the general architecture for the multiple-input MAC operations in Eq.~\ref{eq:mul_mac_op},   which serves as a fundamental building block for DSP and AI computing. Each multiplier takes one input in RG format and another in TB format, with the result encoded in TB format for subsequent propagation and computations.

\begin{figure}[htbp!] \centering
  \includegraphics[width=.8\columnwidth]{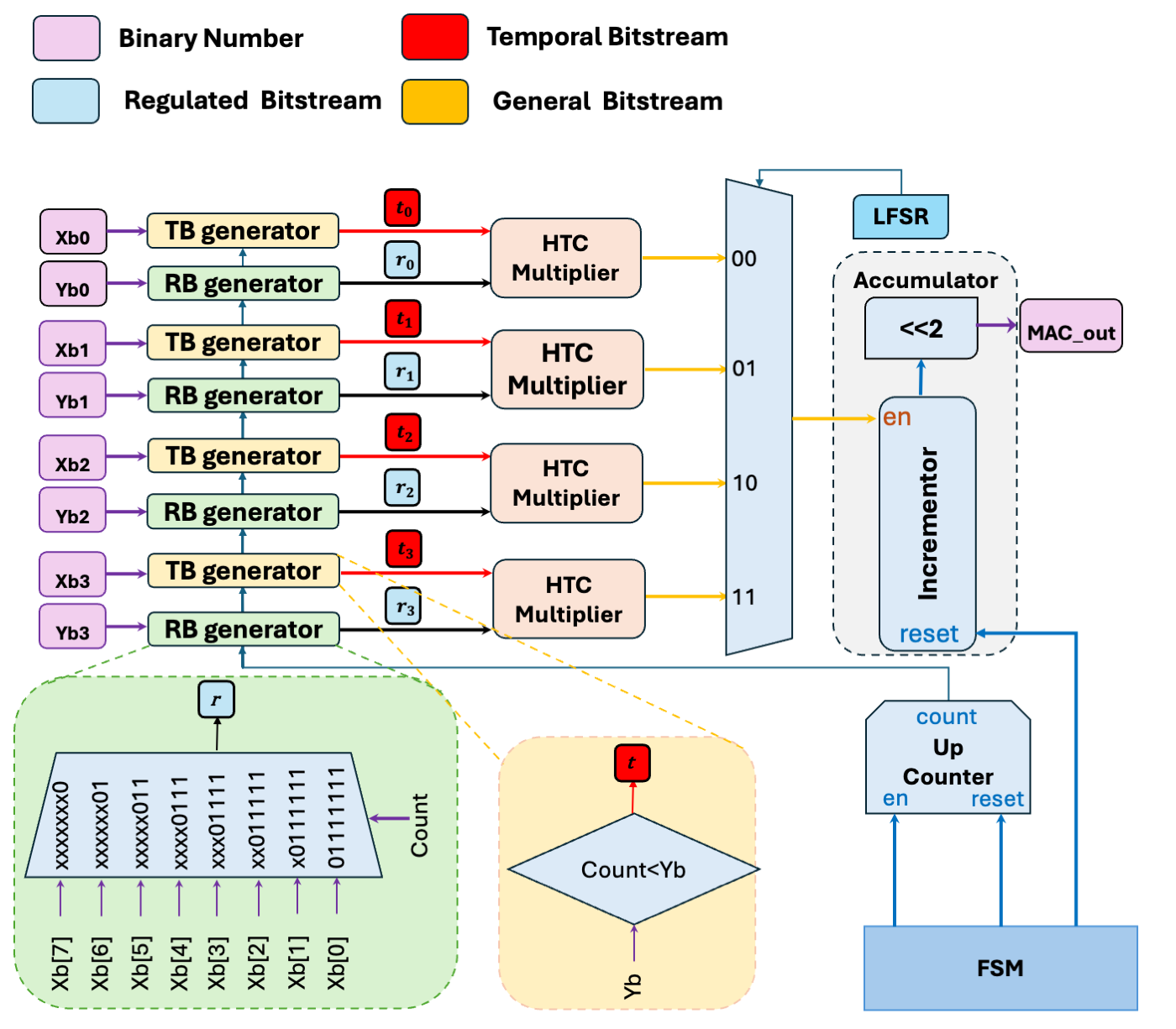}
  \caption{Proposed 4-input HTC MAC design}
  \label{fig:SC_MAC}
\end{figure}

In the following, we present a specific MAC design with \( N = 4 \). The core functional components of the 4-point MAC unit, as illustrated in Fig.~\ref{fig:SC_MAC}, consist of four HTC multipliers and one multiplexer for scaled addition of bitstreams. Each HTC multiplier comprises an AND/XNOR gate for unipolar/bipolar multiplication. This multiplier gate processes one input bitstream from a temporal bitstream (TB) generator ($ti$) and another input bitstream from a regulated bitstream (RB) generator ($ri$). Both the RB and the TB of all multiplicands ($Xbi/Ybi$ in the binary domain) are generated with a single up-counter. 
The output bitstream of each HTC multiplier is directed to an HTC adder that performs scaled addition using a \(4 \times 1\) multiplexer unit. The selector stream for the multiplexer is generated with a linear feedback shift register (LFSR). Furthermore, the same counter is also used to accumulate the bitstream after the scaled addition, either to regenerate the binary number or to generate the downstream temporal bitstream (TB) for the next stage of HTC computation. This HTC accumulator can be implemented with an incrementer and a 2-bit left-shifter. The incrementer is activated by the output bitstream from the MUX gate. The output of the accumulator represents the binary result of the MAC operation.

In this paper, we will use the proposed design of HTC MAC in the two accelerator applications: Finite Impulse Response (FIR) and Discrete Cosine Transform (DCT). Both accelerator operations basically can be expressed in Eq.~\eqref{eq:mul_mac_op}. 

%% file: Experimental_Result.tex
\section{Experimental results and discussions}
\label{sec:results}

In this section, we present the experimental results. First, we discuss the hardware implementation details of the proposed HTC method. Next, we compare the HTC design with two relevant design methods, focusing on single MAC design and two hardware accelerator designs.

\subsection{Hardware implementation details}


For hardware-level evaluation, we implemented the proposed HTC MAC in Verilog hardware description language. The design was synthesized using Synopsys Design Compiler (DC) with the EDK 32nm standard library, which comes with Synopsys University Software Program.  We estimated the area, power, and latency of the HTC MAC using DC tools. For the HTC implementation, the hardware includes all the data format conversions from binary to the temporal bitstram (TB) for coefficients and output data and from binary to regulated bitstream (RG) as well. 


For comparison, we considered two relevant works. The first is the latest deterministic SC design, the Unary MAC~\cite{Schober:IEEE'21}, which is highly relevant as it performs all SC computing in bit-stream formats across all computing steps. The second is the latest CBSC MAC design~\cite{Yu:DAC'21}, which performs addition in binary instead of bit-stream format. All baseline models were implemented using the 32nm EDK standard library and synthesized under the same design constraints in Synopsys Design Compiler for area and power estimation.


For the accuracy comparison of MAC designs, we tested each design with one million random vectors and recorded the results. We report the root mean squared error (RMSE) and standard deviation of error (SDE) for all test vectors. To further evaluate the performance of the proposed method, we also implemented two hardware accelerators for Finite Impulse Response (FIR) and Discrete Cosine Transform (DCT) based on the three methods.


\subsection{Performance comparison for a MAC unit design}
\label{sec:area_power_accuracy_for_mac}

In this section, we compare the proposed HTC MAC design with the Unary MAC design~\cite{Schober:IEEE'21} and the CBSC MAC design~\cite{Yu:DAC'21} in terms of power, area, and latency. For our implementation, each MAC unit computes the dot product between two vectors. Each vector contains four numbers, and each number is an 8-bit unipolar value.



\begin{table}[ht!]\centering
\caption{Comparison for area, power and latency for a MAC unit - (unipolar 8 bit binary number)}\label{Table:Area_power_latency}
\scriptsize
 \begin{tabularx}{\columnwidth}{X*{6}{>{\centering\arraybackslash}X}}
\toprule
\textbf{MAC design}  &\textbf{Area ($\mu m^2$)} 
&\textbf{Power ($\mu W$)} &\textbf{Latency ({\it ns})}&\textbf{RMSE (\%)}&\textbf{SDE (\%)}\\
\midrule
{ CBSC MAC~\cite{Yu:DAC'22}}  & 1476.46 & 56.69 &  2560  & 0.65 &0.42\\
\midrule
{ Unary MAC~\cite{Schober:IEEE'21}}  & 249110.23 & 3085.00 & 870400  & 40.64 & 8.58\\ 
\midrule
{ New HTC MAC }  & 736.95 & 31.07 & 2560 & 6.96 &4.46\\
\bottomrule
\end{tabularx}
\end{table}


The comparison results are presented in Table~\ref{Table:Area_power_latency}. As shown in the table, the proposed HTC MAC delivers significantly smaller power and area footprints compared to the Unary MAC design. It is also orders of magnitude faster than the Unary design. The reason for this is that the Unary MAC involves repetition of input bit streams for the least common multiple (LCM) $(n1, n2)$ times for multiplication, where $n1$ and $n2$ are the bit-stream sizes of the two input data. It also inserts a unique delay for each summand before performing addition using an OR gate, as described in Section ~\ref{sec:review_of_sc}. Consequently, the Unary MAC incurs very large area and latency costs, and thus higher power consumption.

In theory, the Unary MAC can provide the same accuracy as binary designs within a certain input interval range, where it has the maximum allowed number of ones in the bit streams. However, this interval range may not be sufficient for practical computing, given the arbitrary data ranges involved, and errors can still occur. Our results show that the HTC MAC is more accurate than the Unary MAC in our examples, with RMSE and SDE improved by $48.80\%$ and $45.93\%$, respectively, compared to the Unary MAC.


For comparison with the CBSC MAC, the HTC MAC design improves power consumption by $45.19\%$, which is quite significant considering that CBSC is already very power-efficient with less area used. The HTC MAC utilizes a single common up-counter for both the FSM-based regulated bitstream generator (bitstream X) and the counter-based temporal bitstream (bitstream Y) generator in all multiplier units. This optimization reduces the overall area cost of MAC of the same size and decreases dynamic power consumption in the flip-flops of counter-based FSMs and accumulator units. Furthermore, the HTC MAC reduces the area costs by $50.13\%$ compared to the CBSC MAC. However, in terms of accuracy, the HTC MAC is less accurate than the CBSC design. This is expected, as CBSC performs exact addition using binary circuits, while the HTC MAC performs approximate scaled addition.





\subsection{Application I: Finite Impulse Response}

We further implemented a Gaussian blur filter with 6-tap finite impulse response (FIR) filter designs based on the three approaches. The filters' effectiveness was rigorously assessed by processing five distinct images from the USC-SIPI image database~\cite{USC-SIPI}, selected for their varying textures and complexities. To ensure compatibility with the stochastic computational model, the FIR coefficients were quantized into 8-bit unipolar fractional binary numbers. The quality of the filtered images was quantitatively evaluated based on root mean squared error (RMSE) and peak signal-to-noise ratio (PSNR) in dB. 
Table \ref{Table:FIR} presents the PSNR and RMSE of five images blurred with FIR filters implemented using HTC MAC, Unary MAC, and CBSC MAC. The table also reports the hardware costs in terms of area and power consumption for each respective MAC-based filter.

\begin{figure}[htbp]
   \vspace{-0.1in}
    \centering
    \begin{subfigure}{0.4\columnwidth}
    \includegraphics[width=.99\columnwidth]{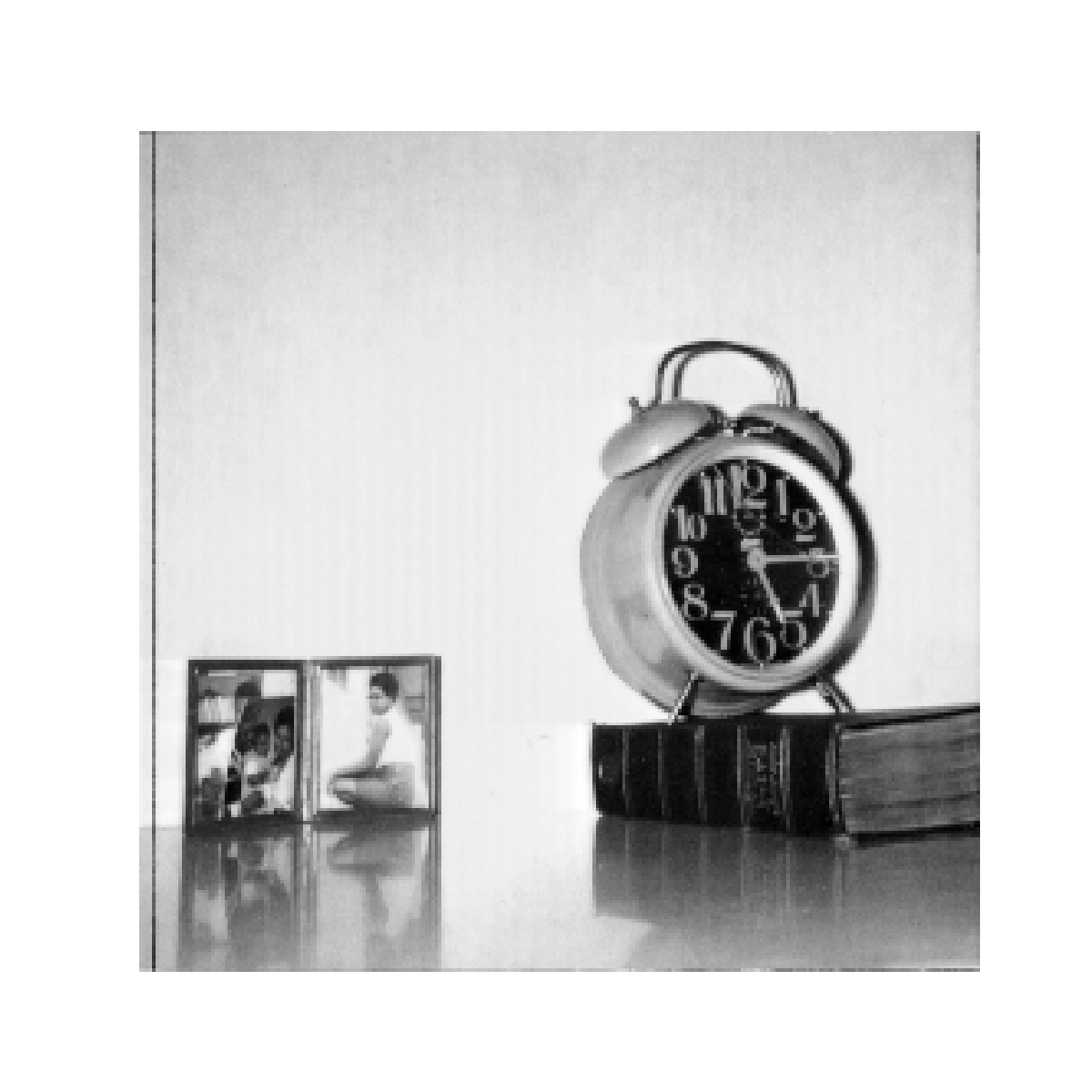}
     \caption{}
     \label{fig:clock_original}
    \end{subfigure}
    \begin{subfigure}{0.4\columnwidth}
    \includegraphics[width=0.99\columnwidth,]{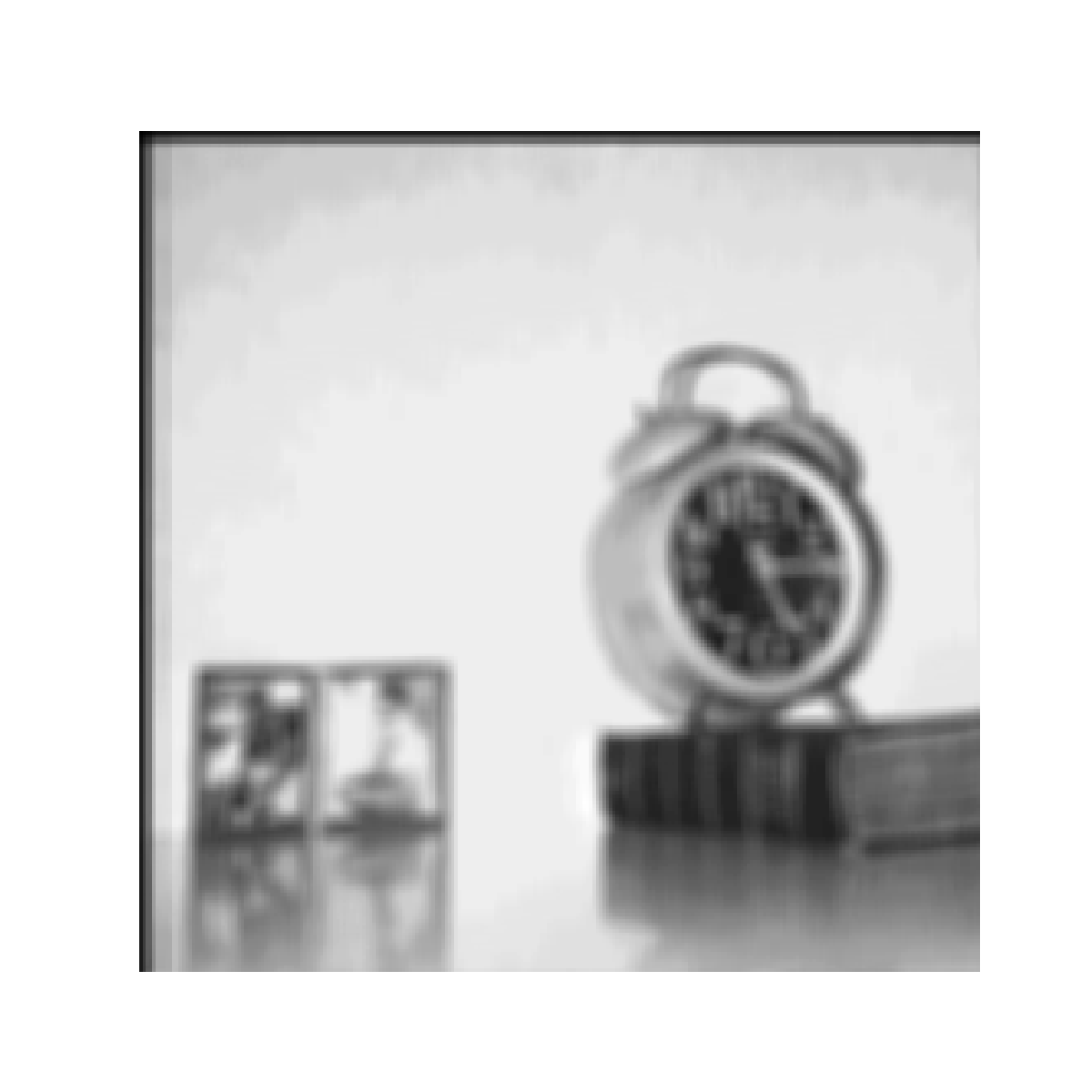}
     \caption{}
     \label{fig:blur_clock_CBSC}
    \end{subfigure}\\
    \begin{subfigure}{0.4\columnwidth}
    \includegraphics[width=0.99\columnwidth,]{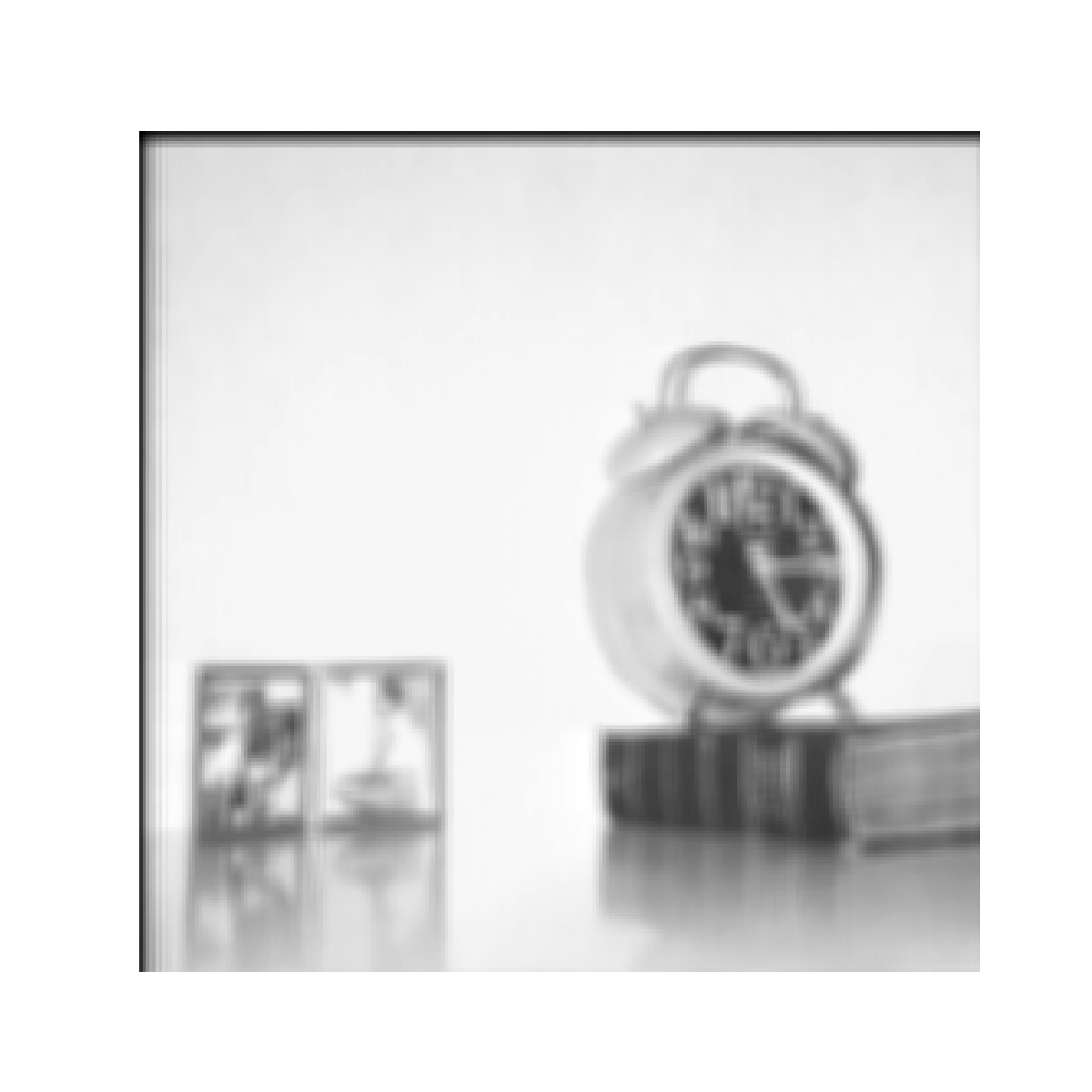}
     \caption{}
     \label{fig:blur_clock_unary}
    \end{subfigure}
    \begin{subfigure}{0.4\columnwidth}
    \includegraphics[width=0.99\columnwidth]{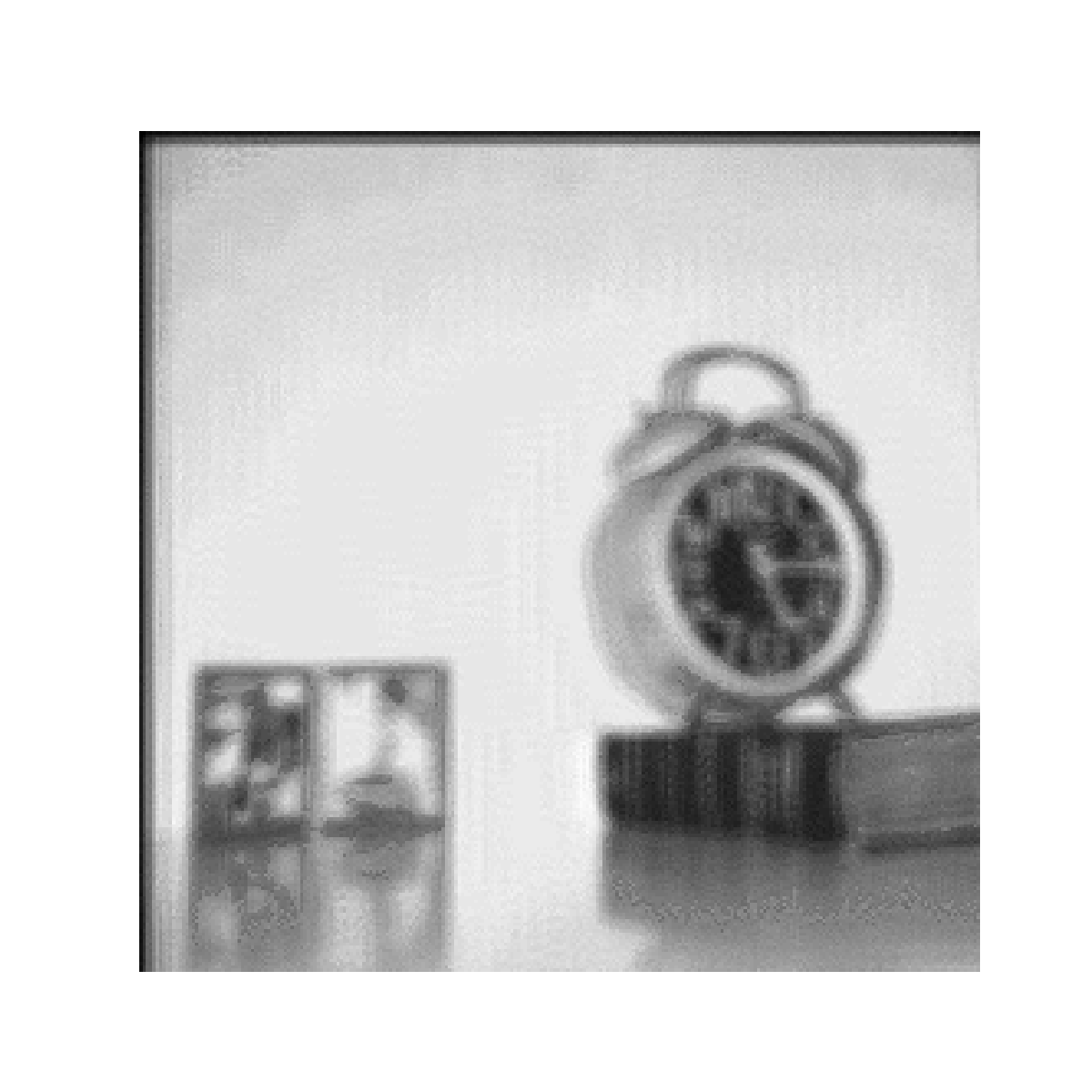}
     \caption{}
     \label{fig:blur_clock_HTC}
    \end{subfigure}
    \\
    \caption{The effect of applying Gaussian blurring filter on (a) original image where the 6-tap unipolar filter is implemented with (b) CBSC MAC, (c) Unary MAC, (d) HTC MAC }
    \label{fig:FIR_clock}
    \vspace{-.25in}
\end{figure}

From the table, we can see that both HTC and CBSC designs significantly outperform the Unary design across all metrics. 
Furthermore, the HTC design delivers PSNR and RMSE comparable to the CBSC design while 
consuming $45.2\%$ less power and occupying  $50.13\%$ less area than its CBSC MAC-based counterpart. These benefits stem from the optimizations highlighted in Section ~\ref{sec:area_power_accuracy_for_mac}. 
Fig.~\ref{fig:FIR_clock} shows the resulting images (the "clock" image) from the three implementations. The results from the HTC (d) and CBSC (b) implementations are quite similar, with notable visual differences compared to the results from the Unary (c) implementation.
 


\begin{table}[ht!]\centering
\caption{Performance comparison for 6-tap FIR-filter}\label{Table:FIR}
\scriptsize
\begin{tabularx}{\columnwidth}{X*{2}{>{\centering\arraybackslash}X}|X*{1}{>{\centering\arraybackslash}X}|X*{1}{>{\centering\arraybackslash}X}}\toprule

\textbf{}& \multicolumn{2}{c|}{\textbf{CBSC MAC~\cite{Yu:DAC'21}}}& \multicolumn{2}{c|}{\textbf{Unary MAC~\cite{Schober:IEEE'21}}}&\multicolumn{2}{c}{\textbf{HTC MAC}} \\ \cmidrule{2-7}
\textbf{Image}&\textbf{PSNR (dB)} &\textbf{RMSE} &\textbf{PSNR (dB)} &\textbf{RMSE} &\textbf{PSNR (dB)} &\textbf{RMSE}
\\\midrule
Boat&20.20&0.10&9.75 &0.325 &20.00 &0.10 \\
Man&21.75&0.08&12.14 &0.247 &21.31 &0.09\\
Couple&20.59&0.09&10.39 &0.301 &20.08 &0.10\\
Bridge&18.67&0.12&10.30 &0.305&18.67 &0.12\\
Clock&17.64&0.13&6.02 &0.499&17.61 &0.13\\
\bottomrule\bottomrule\\
\textbf{Hardware cost}&\textbf{Area ($\mu m^2$)} & \textbf{Power ($\mu W$)} & \textbf{Area   ($\mu m^2$)} & \textbf{Power ($\mu W$)} & \textbf{Area   ($\mu m^2$)} & \textbf{Power ($\mu W$)}\\\midrule
&2091.36 & 62.61 & 8.2$\times10^{5}$ & 10997 & 1216.21 & 39.96\\
\bottomrule
\end{tabularx}
\end{table}


\subsection{Application II: Discrete Cosine Transform}
 We further evaluated our proposed bipolar HTC approach for a widely used image compression tool, the Discrete Cosine Transform (DCT) where the coefficients can be negative. We implemented 8-point DCT accelerators with 8 coefficients using the proposed HTC MAC and CBSC MAC using bipolar encoding. This example compare the performances of the proposed HTC design in bipolar encoding designs. In this application, we did not implement the Unary design as it only works for unipolar data.

\begin{figure}[h]
   \vspace{-0.1in}
    \centering
    \begin{subfigure}{0.32\columnwidth}
    \hspace{-.1in}
    \includegraphics[width=0.99\textwidth]{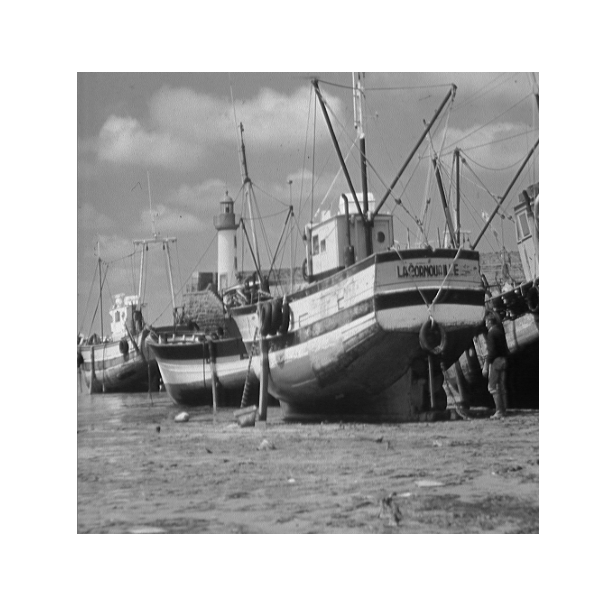}
     \caption{}
     \label{fig:boat_original}
     \hspace{-.1in}
    \end{subfigure}
\begin{subfigure}{0.32\columnwidth}
    \hspace{-.1in}
    \includegraphics[width=0.99\textwidth]{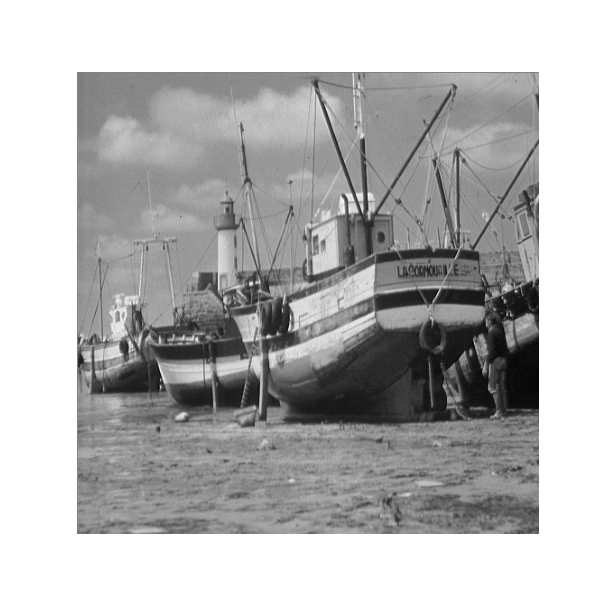}
     \caption{}
     \label{fig:boat_conventional}
     \hspace{-.1in}
    \end{subfigure}
    \begin{subfigure}{0.32\columnwidth}
    \hspace{-.1in}
    \includegraphics[width=0.99\textwidth]{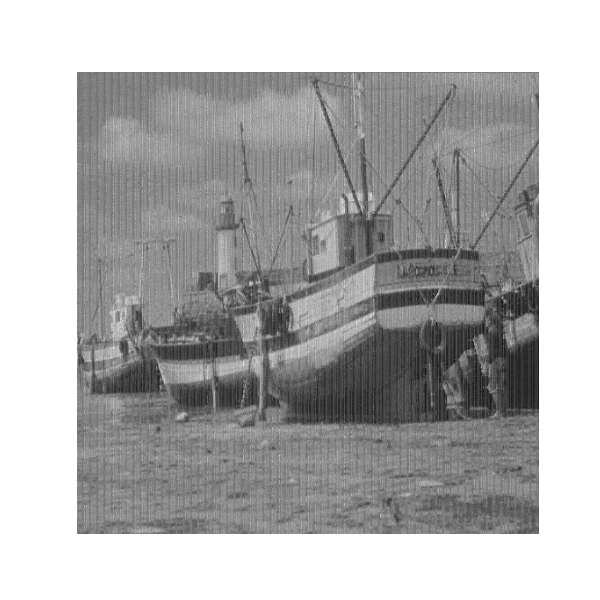}
     \caption{}
     \label{fig:boat_HTC}
     \hspace{-.1in}
    \end{subfigure}
    \\
    \caption{The discrete cosine filter applied to (a) original image of a fishing boat where filter MAC is implemented with (b) CBSC MAC, (c) HTC MAC}
    \label{fig:DCT_boat}
\end{figure}


We first transformed five images of varying resolutions from the USC-SIPI-miscellaneous dataset~\cite{USC-SIPI}. All images, along with the 8-point DCT-II coefficients, were quantized to 8-bit signed fractional binary numbers before applying the filter accelerator to the images. The transformed images were then filtered back to the original domain using an 8-point inverse Discrete Cosine Transform (IDCT) filter. 
The quality of the transformed images was evaluated by computing the peak signal-to-noise ratio (PSNR) in dB and the root mean square error (MSE) of the image transformed back to the original form by the IDCT filter.


Table~\ref{Table:DCT} presents the error metrics and hardware implementation costs of the two designs. The CBSC MAC was implemented with a bipolar CBSC multiplier architecture for multiplying two signed numbers, with exact binary adders used to sum the multiplication results in 2's complement format. The DCT filter was implemented using a 4x1 CBSC MAC, which computes 4 CBSC multiplications and 3 binary additions in this architecture.


The HTC MAC-based DCT filter accelerator  was implemented using the bipolar HTC architecture discussed in Section~\ref{sec:architecture}. It incorporates four bipolar HTC multipliers and one scaled adder to sum the results of the four multiplications. Two four-input HTC MACs are utilized to implement the 8-point approximate DCT filter. All designs were synthesized from the Verilog model using Synopsys Design Compiler with a 32nm technology package.
As depicted in Table~\ref{Table:DCT}, the HTC MAC-based DCT filter achieves up to $23.34\%$ power savings compared to CBSC-based DCT filters. Additionally, the  HTC-based DCT filter uses $18.20\%$ less area than its CBSC-based counterpart .


The PSNR and MSE of five images transformed using the CBSC-based DCT filter and HTC-based DCT filter are also presented in Table~\ref{Table:DCT}. As expected, the CBSC-based design achieves slightly better results.
Fig.~\ref{fig:DCT_boat} illustrates the impact of different MAC-based designs on the DCT transformation of the "boat" image. The proposed energy-efficient HTC MAC-based DCT filter retains the quality of the original image with a PSNR range of approximately 20dB, which is in general, sufficient for such operations. 

\begin{table}[htp!]\centering
\caption{Performance comparison for 8-point DCT}\label{Table:DCT}
\scriptsize
 \begin{tabularx}{.9\columnwidth}{X*{2}{>{\centering\arraybackslash}X}|X*{1}{>{\centering\arraybackslash}X}}\toprule
\textbf{}&\multicolumn{2}{c}{\textbf{CBSC MAC}}&\multicolumn{2}{c}{\textbf{HTC MAC}} \\ \cmidrule{2-5}
\textbf{Image} &\textbf{PSNR (dB)} &\textbf{RMSE} &\textbf{PSNR   (dB)}&\textbf{RMSE}\\\midrule
Boat&38.96&2.89&22.19&19.89\\
Man&37.09&2.69&17.89 &32.63\\
Couple&31.99&6.43&21.78 &20.84\\
Bridge&37.06&3.59&21.39 &21.80\\
Clock&30.95&7.25&21.25 &21.72\\
\bottomrule\bottomrule\\
\textbf{Hardware cost}&\textbf{Area   ($\mu m^2$)} & \textbf{Power ($\mu W$)} & \textbf{Area   ($\mu m^2$)} & \textbf{Power ($\mu W$)} \\\midrule
  & 2532.71 & 81.64& 2071.54 & 62.59\\
\bottomrule
\end{tabularx}
\end{table}

%% file: Conclusion.tex
\section{Conclusion}
\label{sec:conclusion}
In this paper, we have developed a new hybrid temporal computing (HTC), which overcomes these limitations by encoding signals in both temporal and pulse rate formats for multiplication and in temporal format for propagation. We showed how the basic arithmetic units like multiplier and accumulator (MAC) are implemented and demonstrated HTC on two applications: digital Finite Impulse Response (FIR) filter and DCT/iDCT engine design.
Experimental results show that the HTC MAC uses only a small marginal power and area footprint compared to the Unary MAC design. It is also orders of magnitude faster than the Unary design. Compared to the CBSC MAC, it reduces power consumption by $45.2\%$ and area footprint by $50.13\%$. For the FIR filter design, HTC designs again significantly outperform the Unary design across all metrics. Compared to CBSC design, the HTC MAC-based FIR filter reduces power consumption by $36.61\%$ and area cost by $45.85\%$.  The HTC-based DCT filter retains the quality of the original image with a decent PSNR, while consuming $23.34\%$ less power and occupying $18.20\%$ less area than the state-of-the-art CBSC MAC-based DCT filter.
